\documentclass[
groupaddress,aps,preprintnumbers,amsmath,amssymb,prd,nofootinbib,preprint]{revtex4-1}

\usepackage[dvipdfm]{graphicx}
\usepackage{epstopdf}
\usepackage{dcolumn}
\usepackage{bm}
\usepackage{hyperref}
\usepackage{color}
\usepackage{setspace}

\begin{document}


\def\a{\alpha}
\def\b{\beta}
\def\c{\varepsilon}
\def\d{\delta}
\def\e{\epsilon}
\def\f{\phi}
\def\g{\gamma}
\def\h{\theta}
\def\k{\kappa}
\def\l{\lambda}
\def\m{\mu}
\def\n{\nu}
\def\p{\psi}
\def\q{\partial}
\def\r{\rho}
\def\s{\sigma}
\def\t{\tau}
\def\u{\upsilon}
\def\v{\varphi}
\def\w{\omega}
\def\x{\xi}
\def\y{\eta}
\def\z{\zeta}
\def\D{\Delta}
\def\G{\Gamma}
\def\H{\Theta}
\def\L{\Lambda}
\def\F{\Phi}
\def\P{\Psi}
\def\S{\Sigma}

\def\o{\over}
\def\beq{\begin{eqnarray}}
\def\eeq{\end{eqnarray}}
\newcommand{\gsim}{ \mathop{}_{\textstyle \sim}^{\textstyle >} }
\newcommand{\lsim}{ \mathop{}_{\textstyle \sim}^{\textstyle <} }
\newcommand{\vev}[1]{ \left\langle {#1} \right\rangle }
\newcommand{\bra}[1]{ \langle {#1} | }
\newcommand{\ket}[1]{ | {#1} \rangle }
\newcommand{\EV}{ {\rm eV} }
\newcommand{\KEV}{ {\rm keV} }
\newcommand{\MEV}{ {\rm MeV} }
\newcommand{\GEV}{ {\rm GeV} }
\newcommand{\TEV}{ {\rm TeV} }
\newcommand{\1}{\mbox{1}\hspace{-0.25em}\mbox{l}}
\def\diag{\mathop{\rm diag}\nolimits}
\def\Spin{\mathop{\rm Spin}}
\def\SO{\mathop{\rm SO}}
\def\O{\mathop{\rm O}}
\def\SU{\mathop{\rm SU}}
\def\U{\mathop{\rm U}}
\def\Sp{\mathop{\rm Sp}}
\def\SL{\mathop{\rm SL}}
\def\tr{\mathop{\rm tr}}

\def\IJMP{Int.~J.~Mod.~Phys. }
\def\MPL{Mod.~Phys.~Lett. }
\def\NP{Nucl.~Phys. }
\def\PL{Phys.~Lett. }
\def\PR{Phys.~Rev. }
\def\PRL{Phys.~Rev.~Lett. }
\def\PTP{Prog.~Theor.~Phys. }
\def\ZP{Z.~Phys. }

\def\dd{\mathrm{d}}
\def\ff{\mathrm{f}}
\def\BH{{\rm BH}}
\def\inf{{\rm inf}}
\def\ev{{\rm evap}}
\def\eq{{\rm eq}}
\def\SM{{\rm sm}}
\def\Mpl{M_{\rm Pl}}
\def\GeV{{\rm GeV}}
\newcommand{\Red}[1]{\textcolor{red}{#1}}

\def\mDM{m_{\rm DM}}
\def\mphi{m_{\phi}}
\def\TeV{{\rm TeV}}
\def\Gphi{\Gamma_\phi}
\def\TR{T_{\rm RH}}
\def\Br{{\rm Br}}
\def\DM{{\rm DM}}
\def\Eth{E_{\rm th}}
\newcommand{\lmk}{\left(}  
\newcommand{\rmk}{\right)}
\newcommand{\lkk}{\left[}  
\newcommand{\rkk}{\right]}
\newcommand{\lhk}{\left \{ }  
\newcommand{\rhk}{\right \} }
\newcommand{\del}{\partial}  
\newcommand{\la}{\left\langle} 
\newcommand{\ra}{\right\rangle}


\title{
Pure Gravity Mediation and Chaotic Inflation in Supergravity
}

\author{Keisuke Harigaya}
\author{Tsutomu T.~Yanagida}
\affiliation{Kavli IPMU (WPI), TODIAS, University of Tokyo, Kashiwa, 277-8583, Japan}
\begin{abstract}
We investigate compatibility of the pure gravity mediation (or the minimal split supersymmetry)
with chaotic inflation models in supergravity.
We find that an approximate $Z_2$ parity of the inflaton is useful to suppress gravitino production from thermal bath and to obtain consistent inflation dynamics.
We discuss production of the lightest supersymmetric particle through the decay of the inflaton with the approximate $Z_2$ symmetry
and find that a large gravitino mass is favored to avoid the overproduction of the lightest supersymmetric particle,
while a lower gravitino mass requires tuning of parameters.
This may explain why the nature has chosen the gravitino mass of $O(100)$ TeV
rather than $O(100)$ GeV.
\end{abstract}

\date{\today}
\maketitle
\preprint{IPMU 14-0154}

\section{Introduction}

High scale supersymmetry (SUSY) with the gravitino mass $m_{3/2}=O(100)$ TeV  is one of the most interesting models beyond the standard model.
It not only explains the observed Higgs boson mass $m_h\simeq 126$ GeV~\cite{Aad:2012tfa,Chatrchyan:2012ufa} by stop and top-loop radiative corrections~\cite{Okada:1990gg,Ellis:1990nz,Haber:1990aw},
but also it is free from serious phenomenological and gravitino problems thanks to large sfermion and gravitino masses, $m_{\rm sfermion} \simeq m_{3/2}=O(100)$ TeV.
Among high scale SUSY models, the pure gravity mediation (PGM)~\cite{Ibe:2006de,Ibe:2011aa,Ibe:2012hu}
is a particularly attractive scenario,
for we do not need to introduce the Polonyi field to generate gaugino masses~\cite{Giudice:1998xp,Randall:1998uk} and the SUSY invariant mass (so-called $\mu$) term of higgs multiplets~\cite{Inoue:1991rk,Casas:1992mk} in the minimal SUSY standard model (MSSM).
Thus, the model is completely free from the cosmological Polonyi problem \cite{Coughlan:1983ci,Ibe:2006am}
(see also  the minimal split SUSY \cite{ArkaniHamed:2012gw} whose basic structure is identical to the PGM.).%
\footnote{
High scale SUSY models are also discussed in Refs~\cite{Hall:2011jd,Arvanitaki:2012ps}.
In Ref.~\cite{Hall:2011jd}, a mediation scale other than the Planck scale is
introduced to generate soft scalar masses, and hence soft masses have a broader range than
in the case of the PGM.
In Ref~\cite{Arvanitaki:2012ps}, the Polonyi field is introduced to generate the $\mu$ term, and hence it is essentially
different from the PGM.
}

On the other hand, the chaotic inflation~\cite{Linde:1983gd} is one of the most attractive cosmic inflation scenarios~\cite{Guth:1980zm,Kazanas:1980tx}.
It is free from the initial condition problem~\cite{Linde:2005ht}.
That is, inflation takes place for generic initial conditions of the inflaton field and the space-time.
The chaotic inflation has been successfully realized in the supergravity (SUGRA)~\cite{Kawasaki:2000yn}.

In this paper, we investigate compatibility of the PGM with the chaotic inflation.
We show that in the PGM, the inflaton should have a $Z_2$ odd parity to suppress the reheating temperature, avoiding the gravitino overproduction from thermal bath~\cite{Weinberg:1982zq,Nanopoulos:1983up,Ellis:1984eq,Kawasaki:1994af}.
We also show that the $Z_2$ symmetry is helpful for the inflaton to have consistent dynamics without tuning of parameters in the inflaton sector.

In order for the inflaton to decay, we argue that the $Z_2$ symmetry is softly broken by a small amount.
We discuss the reheating process assuming the small breaking of the $Z_2$ symmetry, with paying attention to the gravitino overproduction problem.
It is known that the inflaton in general decays into gravitinos, which leads to the overproduction of the lightest SUSY particle (LSP)~\cite{Kawasaki:2006gs,Asaka:2006bv,Dine:2006ii,Endo:2006tf,Endo:2006qk,Endo:2007ih}.
We consider that the LSP is stable and a candidate for dark matter (DM) in the universe.
We discuss how the overproduction of the LSP can be avoided.
Assuming that leptogenesis~\cite{Fukugita:1986hr} (for a review, see Ref.~\cite{Buchmuller:2005eh}) is responsible for the origin of the baryon asymmetry in the universe,
we show that our solution to the above problem suggests a gravitino mass far larger than the electroweak scale, $m_{3/2}\gsim O(100)$~TeV, while
fine tuning of parameters in the SUSY breaking sector and the MSSM sector is required for
a smaller gravitino mass.
We note that we do not use any constraints from the successful Big Bang Nucleosynthesis (BBN) to derive the natural lower bound on the gravitino mass.

\begin{figure}[tb]
\begin{center}
  \includegraphics[width=.6\linewidth]{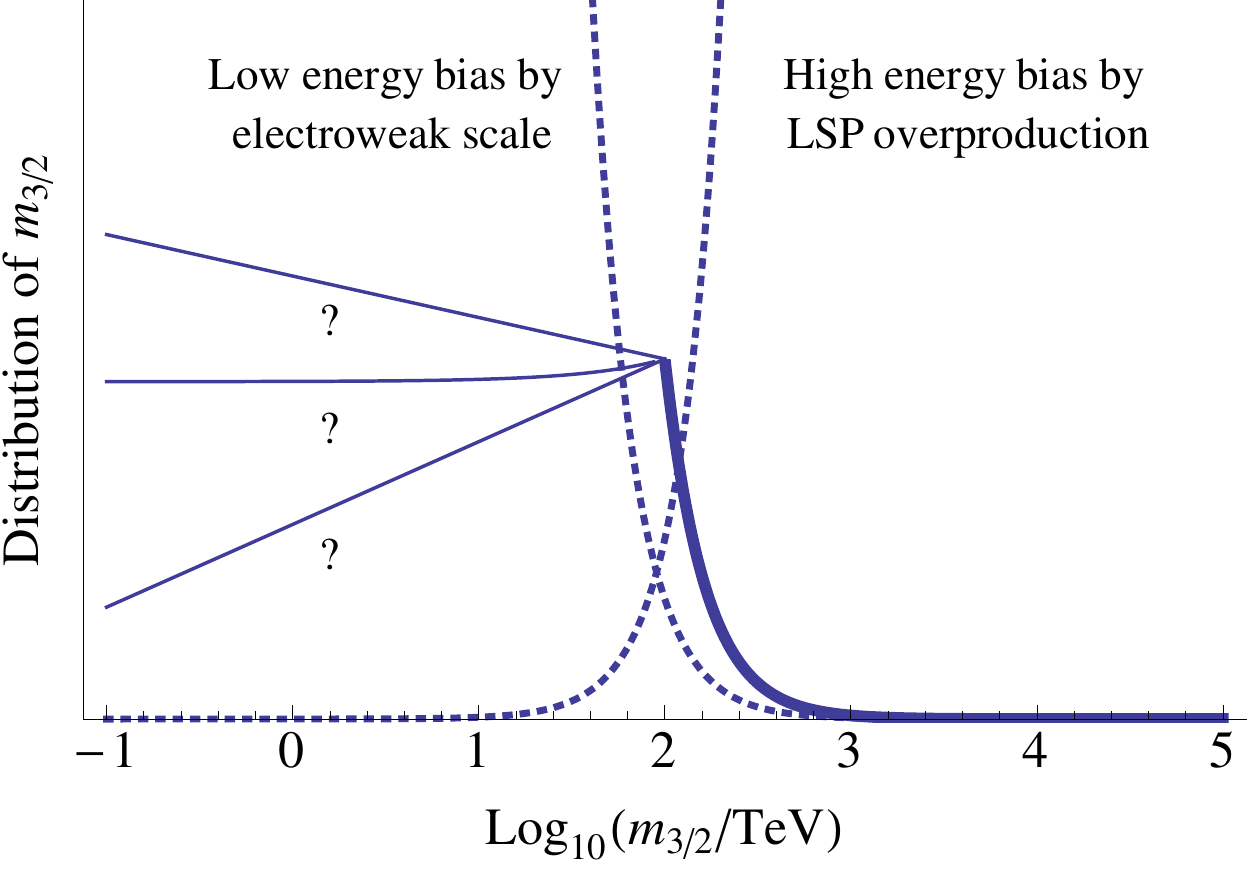}
 \end{center}
\caption{\sl \small
A sketch on possible distributions of the gravitino mass. 
}
\label{fig:bias}
\end{figure}

This may answer to a fundamental question for the high scale SUSY;
why the nature chooses the high scale SUSY with $m_{3/2}= O(100)$ TeV, but not so-called a natural SUSY with $m_{3/2}= O(100)$ GeV.
The gravitino mass was in fact expected of $O(100)$ GeV before the Large Hadron Collider,
for the electroweak scale is naturally obtained without tuning of parameters in the MSSM when $m_{3/2} = O(100)$~GeV.
In the landscape point of view~\cite{Bousso:2000xa,Kachru:2003aw,Susskind:2003kw,Denef:2004ze}, it seems to be difficult to understand why the nature chooses $m_{3/2}= O(100)$ TeV.
As we show in this paper, the gravitino mass of $O(100)$ GeV requires fine tuning to avoid the LSP overproduction,
otherwise the DM density of the present universe is outside the anthropic window~\cite{Hellerman:2005yi,Tegmark:2005dy}.
Thus, $m_{3/2}= O(100)$ TeV may be as plausible as $m_{3/2}= O(100)$ GeV (see Fig.~\ref{fig:bias} for the schematic picture).

This paper is organized as follows.
In the next section, we review chaotic inflation models in the supergravity and show that the inflaton should have a $Z_2$ odd parity in the PGM.
In Sec.~\ref{sec:gravitino}, we discuss the decay of the inflaton into gravitinos and show how the LSP overproduction can be avoided.
We show that the solution to the LSP overproduction problem favors a gravitino mass far larger than the electroweak scale and smaller gravitino masses require tuning of parameters.
The last section is devoted to discussion and conclusions.

\section{Chaotic inflation model in supergravity}
\label{sec:model}

The chaotic inflation~\cite{Linde:1983gd} is an attractive inflation model, for it is free from the initial condition problem~\cite{Linde:2005ht}:
Inflation takes place for generic initial conditions of the inflaton field and the space-time.
In this section, we review chaotic inflation in SUGRA proposed in Ref.~\cite{Kawasaki:2000yn}.
We first discuss the inflaton dynamics in chaotic inflation model in SUGRA.
Next, we show that the inflaton should have a $Z_2$ odd parity in the PGM.
In order for the inflaton to decay, we assume that the $Z_2$ symmetry is explicitly broken by a small value of a spurious field ${\cal E}$.
Then, we discuss the decay of the inflaton into MSSM fields.

\subsection{Supergravity chaotic inflation model}

In SUGRA, the scalar potential is given by
the Kahler potential $K(\phi^i,\phi^{*\bar{i}})$ and the superpotential $W(\phi^i)$,
where $\phi^i$ and $\phi^{*\bar{i}}$ are chiral multiplets and their conjugate anti-chiral multiplets, respectively.
The scalar potential is given by
\begin{eqnarray}
\label{eq:potential}
V &=&  e^K \left[
K^{\bar{i}i}D_i W D_{\bar{i}}W^* - 3 |W|^2
\right],\nonumber\\
D_i W &\equiv& W_i + K_i W,
\end{eqnarray}
where subscripts $i$ and $\bar{i}$ denote derivatives with respect to $\phi^{i}$ and $\phi^{*\bar{i}}$, respectively.
$K^{\bar{i}i}$ is the inverse of the matrix $K_{i\bar{i}}$.
Here and hereafter, we use a unit of the reduced Planck mass $M_{\rm pl}\simeq 2.4\times 10^{18}$ GeV being unity.

The chaotic inflation requires a large field value of the inflaton during inflation.
With the large field value, the slow-roll inflation seems to be difficult to take place in SUGRA,
because of the exponential factor in the scalar potential, $e^K$.
This problem was naturally solved in Ref.~\cite{Kawasaki:2000yn} by assuming a shift symmetry of the inflaton chiral multiplet $\Phi$,
\begin{eqnarray}
\label{eq:shift}
\Phi \rightarrow \Phi + i C,
\end{eqnarray}
where $C$ is a real number.
The inflaton is identified with the imaginary scalar component of $\Phi$, $\phi\equiv \sqrt{2}{\rm Im}\Phi$.
The exponential factor vanishes for $\phi$, and hence the slow-roll inflation is naturally realized for a large field value of $\phi$.
Note that the chiral multiplet $\Phi$ must have a vanishing $R$ charge to be consistent with the shift symmetry.

The inflaton potential is obtained by breaking the shift symmetry softly in the superpotenial,%
\footnote{For discussion on the shift symmetry breaking in the Kahler potential, see Refs.~\cite{Kallosh:2010ug,Kallosh:2010xz,Li:2013nfa,Harigaya:2014qza}.}
\begin{eqnarray}
W = m X \Phi,
\end{eqnarray}
where $X$ is a chiral multiplet with an $R$ charge of 2.
The explicite breaking of the shift symmetry is expressed by the parameter $m$.
Here, we have eliminated the term allowed by the $R$ symmetry, $W\supset M^2 X$, where $M$ is a constant,
by a redefinition $\Phi \rightarrow \Phi - M^2/m$.

Let us discuss the inflaton dynamics. The Kahler potential consistent with the shift symmetry is given by
\begin{eqnarray}
\label{eq:Kahler inf}
K = c (\Phi + \Phi^\dag) + \frac{1}{2}(\Phi + \Phi^\dag)^2 + X X^\dag + \cdots,
\end{eqnarray}
where $\cdots$ denotes higher dimensional terms, which we neglect for simplicity.
The scalar potential is given by
\begin{eqnarray}
V(\phi,\sigma) = {\rm exp}\left(\sigma^2 + \sqrt{2} c \sigma\right) \frac{1}{2}m^2\left( \phi^2 + \sigma^2\right),
\end{eqnarray}
where $\sigma$ is the real scalar component of $\Phi$, $\sigma \equiv \sqrt{2} {\rm Re} \Phi $.
Since $X$ is stabilized near the origin during inflation by a Hubble induced mass term, we have set $X=0$~\cite{Kawasaki:2000yn}.
For given $\phi \gg 1$, the scalar potential is minimized for $\sigma = - c /\sqrt{2}$.
Thus, the scalar potential of $\phi$ during inflation is given by
\begin{eqnarray}
V_{\rm inf}(\phi) \simeq \frac{1}{2}m_{\rm eff}^2\phi^2, ~~~ m_{\rm eff} \equiv m \times e^{- c^2/4}.
\end{eqnarray}
The observed magnitude of the curvature perturbation, ${\cal P}_\zeta \simeq 2.2 \times 10^{-9}$~\cite{Ade:2013zuv}, determines $m_{\rm eff}$ as
\begin{eqnarray}
m_{\rm eff} \simeq 6.0 \times 10^{-6} = 1.5 \times 10^{13}~{\rm GeV},
\end{eqnarray}
where we have assumed that the number of e-foldings corresponding to the pivot scale of $0.002~{\rm Mpc}^{-1}$ is as large as $50-60$.

After inflation, extremums of the potential is given by
\begin{eqnarray}
\label{eq:minimum}
\frac{\partial V}{\partial \phi} \propto \phi =0,~~
\frac{\partial V}{\partial \sigma} \propto \sigma \left( \sigma^2 + \frac{c}{\sqrt{2}} \sigma +1 \right) + \phi^2 \left(\sigma +\frac{c}{\sqrt{2}} \right) =0.
\end{eqnarray}
For $c^2 < 8$, Eq.~(\ref{eq:minimum}) has a unique solution at the origin.
For $c^2 > 8$, Eq.~(\ref{eq:minimum}) has three solutions for $\sigma$.
One of the solutions, $\sigma = 0$, is the minimum with a vanishing potential and another solution, $\sigma = - c/ (2 \sqrt{2}) -{\rm sgn}(c) \sqrt{c^2/8 -1}$, is a local minimum with a non-vanishing potential.
The other is a local maximum.
Since $\sigma$ is trapped at $\sigma = - c /\sqrt{2}$ for large $\phi$ values, as $\phi$ becomes small, $\sigma$ moves to the local minimum with a non-vanishing potential, which prevents the inflation from ending.
Thus, it is  required that $c^2<8$.

At around the origin, the mass of the inflaton is larger than $m_{\rm eff}$. Since $c^2<8$, the mass of the inflaton at the origin, $m$, is within the range of~%
\footnote{This range is slightly widen by taking account of higher dimensional terms in the Kahler potential.
Even if the mass of the inflaton is as large as $10^{14}$~GeV and hence the decay of the inflaton after inflation produces particles with extremely large momenta, the decay products thermalize soon after their production~\cite{Harigaya:2013vwa}. Thus, the standard estimation of the reheating temperature in the following discussion is valid.
}
\begin{eqnarray}
1.5 \times 10^{13}~{\rm GeV} = m_{\rm eff} \leq m < e^2 m_{\rm eff} = 1.1 \times 10^{14}~{\rm GeV}.
\end{eqnarray}

\subsection{Motivation of a $Z_2$ symmetry}

Let us consider possible couplings of the inflaton to the MSSM particles.
We first note that the field $X$ has an $R$ charge of $2$.
This is mandatory because the inflaton multiplet $\Phi$ must possess a shift symmetry, so its $R$ charge must vanish.
On the other hand, in the PGM, the higgsino Dirac mass term, so called the $\mu$ term, is generated by the tree level coupling of the higgs multiplets to the $R$ symmetry breaking~\cite{Inoue:1991rk,Casas:1992mk}.
This ensures the $\mu$ term to be of the same order as the soft scalar mass term, i.e., the gravitino mass.
This mechanism requires the combination $H_u H_d$, where $H_u$ and $H_d$ are the up and down type higgs multiplets, to have vanishing charges under any symmetry. Therefore, the following superpotential term is not forbidden by the $R$ symmetry,
\begin{eqnarray}
\label{eq:XHH}
W \supset g X H_u H_d,
\end{eqnarray}
where $g$ is a constant.

The inflaton decays into higgs pairs through the coupling in Eq.~(\ref{eq:XHH}). The resultant reheating temperature is
\begin{eqnarray}
T_{\rm RH} = 1.5\times 10^9~{\rm GeV}~\frac{g}{m} \left(\frac{m}{1.5\times 10^{13}~{\rm GeV}}\right)^{1/2}.
\end{eqnarray}
For $g = O(1)$, the reheating temperature is so high that too many gravitinos are produced through thermal scatterings~\cite{Weinberg:1982zq,Nanopoulos:1983up,Ellis:1984eq,Kawasaki:1994af}.
The coupling $g$ must be extremely suppressed~\cite{Kawasaki:2000yn}.%
\footnote{
If $g$ is not suppressed, the $F$ term of $X$ strongly depends on $H_uH_d$. The $H_u H_d$ direction works as a waterfall field in the hybrid inflation~\cite{Linde:1991km}, and thus inflation ends for $|\phi|\gg 1$. This changes the prediction on the spectral index and the tensor fraction. We note that during the waterfall phase, the instability of $H_u$ and $H_d$ grows and the reheating temperature becomes extremely high. 
}
The suppression is easily achieved if $X$ and $\Phi$ are odd under a $Z_2$ symmetry.%
\footnote{The $Z_2$ symmetry is consistent with the shift symmetry given in Eq.~(\ref{eq:shift}).
$g$ can be also suppressed if $H_u H_d$ carries a Peccei-Quinn charge. For the PGM model with the Peccei-Quinn symmetry, see Refs.~\cite{Feldstein:2012bu,Evans:2014hda}.
}
We note that the $Z_2$ symmetry is also helpful to have successful inflaton dynamics.
As we have mentioned in the previous subsection, the superpotential term of $W\supset M^2 X$ is allowed by the $R$ symmetry.
The constant $M$ is expected to be of order one without the $Z_2$ symmetry.
As we shift $\Phi$, $\Phi \rightarrow \Phi - M^2 /m$, to eliminate the superpotential term, a large linear term in the Kahler potential, $c (\Phi + \Phi^\dag)$ in Eq.~(\ref{eq:Kahler inf}), is induced.
However, for inflation to end, the constant $c$ in Eq.~(\ref{eq:Kahler inf}) must be smaller than $\sqrt{8}$, which requires tuning among parameters in the Kahler potential.
We can easily avoid the tuning if we have the $Z_2$ symmetry.

Taking those problems seriously, we assume, throughout this paper, the $Z_2$ symmetry under which $X$ and $\Phi$ are odd.
In order for the inflaton to decay
into the MSSM particles,
we assume that the $Z_2$ symmetry is broken by a small amount,
which we express by a spurious field ${\cal E}$.%
\footnote{Alternatively, the inflaton can decay into MSSM fields if MSSM fields are also charged under the $Z_2$ symmetry~\cite{Kawasaki:2000ws,Harigaya:2014pqa}. We do not consider this possibility in this paper.}
Here, the spurion ${\cal E}$ is odd under the $Z_2$ symmetry and a non-vanishing value of ${\cal E}$ represents the $Z_2$ symmetry breaking.

\subsection{Decay of the inflaton into MSSM fields}

Based on the assumption of the broken $Z_2$ symmetry, we consider the following super and Kahler potential for the inflaton and the MSSM sectors,

\begin{eqnarray}
\label{eq:superKahler}
W &=& X( m \Phi - {\cal E} ) + a_1{\cal E} X H_u H_d + W_{\rm MSSM},\nonumber\\
K &=& X X^\dag + \frac{1}{2}(\Phi + \Phi^\dag)^2 + Q Q^\dag,
\end{eqnarray}
where $W_{\rm MSSM}$ is the superpotential of the MSSM, $Q$ denotes MSSM fields collectively, and $a_1$ is an order one coefficient. We take $m$ to be real without loss of generality.
To be concrete, we have assumed the minimal form of the Kahler potential.
For clarity, we shift $\Phi$ as $\Phi \rightarrow \Phi + {\cal E} / m$. Then the super and the Kahler potential is given by
\begin{eqnarray}
\label{eq:Kahlersuper}
W &=& mX \Phi  + a_1 {\cal E} X H_u H_d + W_{\rm MSSM},\nonumber\\
K &=& X X^\dag + \frac{1}{2}(\Phi + \Phi^\dag)^2 + c(\Phi + \Phi^\dag) +  Q Q^\dag,
\end{eqnarray}
where $c\equiv ({\cal E} + {\cal E}^\dag) / m$ is a real constant.
For a successful inflation, $c$ must be smaller than $\sqrt{8}$, which indicates that $|{\cal E}| < O (m)$.

Let us discuss the decay of the inflaton into MSSM fields.
First, the inflaton decays into higgs pairs through the coupling in the superpotential in Eq.~(\ref{eq:superKahler}) with the width,
\begin{eqnarray}
\label{eq:decay treeMSSM}
\Gamma (\phi \rightarrow H_u H_d) = \frac{1}{4\pi} |a_1{\cal E}|^2 m.
\end{eqnarray}

Second, the inflaton automatically decays through the linear term of the inflaton field in the Kahler potential, if a non-vanishing superpotential of MSSM fields exists~\cite{Endo:2006qk,Endo:2007ih}.
Assuming the presence of right-handed neutrinos with Majorana masses to explain the neutrino mass~\cite{seesaw},
dominant decay modes are provided by
the following superpotential,
\begin{eqnarray}
W = y_t Q_3 \bar{u}_3 H_u + \frac{1}{2}M_N N N,
\end{eqnarray}
where $Q_3$, $\bar{u}_3$ and $N$ are the third-generation quark doublet, the third-generation up-type quark, and a right-handed neutrino, respectively. 
$y_t$ and $M_N$ are the top yukawa coupling and the right-handed neutrino mass, respectively.
For simplicity, we assume that only one right-handed neutrino is lighter than the inflaton.
Decay widths of the inflaton by these interactions are
\begin{eqnarray}
\label{eq:decay superMSSM}
\Gamma(\phi \rightarrow Q_3 \bar{u}_3 H_u) &=& \frac{3}{128\pi^3}c^2 y_t^2 m^3,\nonumber\\
\Gamma(\phi \rightarrow NN) &=& \frac{1}{16\pi}c^2 m M_N^2.
\end{eqnarray}

Third, the inflaton couples with gauge multiplets through radiative corrections~\cite{Endo:2007ih}.
Radiative corrections induce couplings of the inflaton in kinetic functions,%
\footnote{When one moves on to the Einstein frame and canonicalizes fields, one encounters inflaton-dependent chiral rotations of fermions fields.
Thus, in the Einstein frame with canonical normalization for matter and gauge fields,
the shift symmetry also involves chiral rotations of fermions fields, which is anomalous.
The coupling in Eq.~(\ref{eq:anomaly}) can be understood as the counter term for the anomaly.
}
\begin{eqnarray}
\label{eq:anomaly}
\left[\frac{1}{g^2} + i\frac{\theta_{\rm YM}}{8\pi^2} + \frac{1}{16\pi^2}c \Phi (T_{\rm G}-T_M) \right] W^\alpha W_\alpha,
\end{eqnarray}
where $g$, $\theta_{\rm YM}$ and $W^\alpha$ are the gauge coupling constant, the theta angle, and the field strength superfield, respectively.
$T_G$ is the Dynkin index of the adjoint representation and $T_M$ is the total Dynkin index of matter fields.
The decay width of the inflaton into the gauge multiplet $V$ by the gauge kinetic function is given by
\begin{eqnarray}
\label{eq:decay anomaly}
\Gamma(\phi \rightarrow V V) = \frac{\alpha^2}{256\pi^3} N_G (T_G-T_M)^2 c^2 m^3,
\end{eqnarray}
where $\alpha = g^2 /4\pi$ and $N_G$ is the number of the generator of the gauge symmetry.
Due to the suppression by a one-loop factor, this decay mode is sub-dominant in the MSSM.
As we will see, however, this decay mode plays an important role in considering the
decay of the inflaton into the SUSY breaking sector in Sec.~\ref{sec:gravitino}.

\begin{figure}[tb]
\begin{center}
  \includegraphics[width=.6\linewidth]{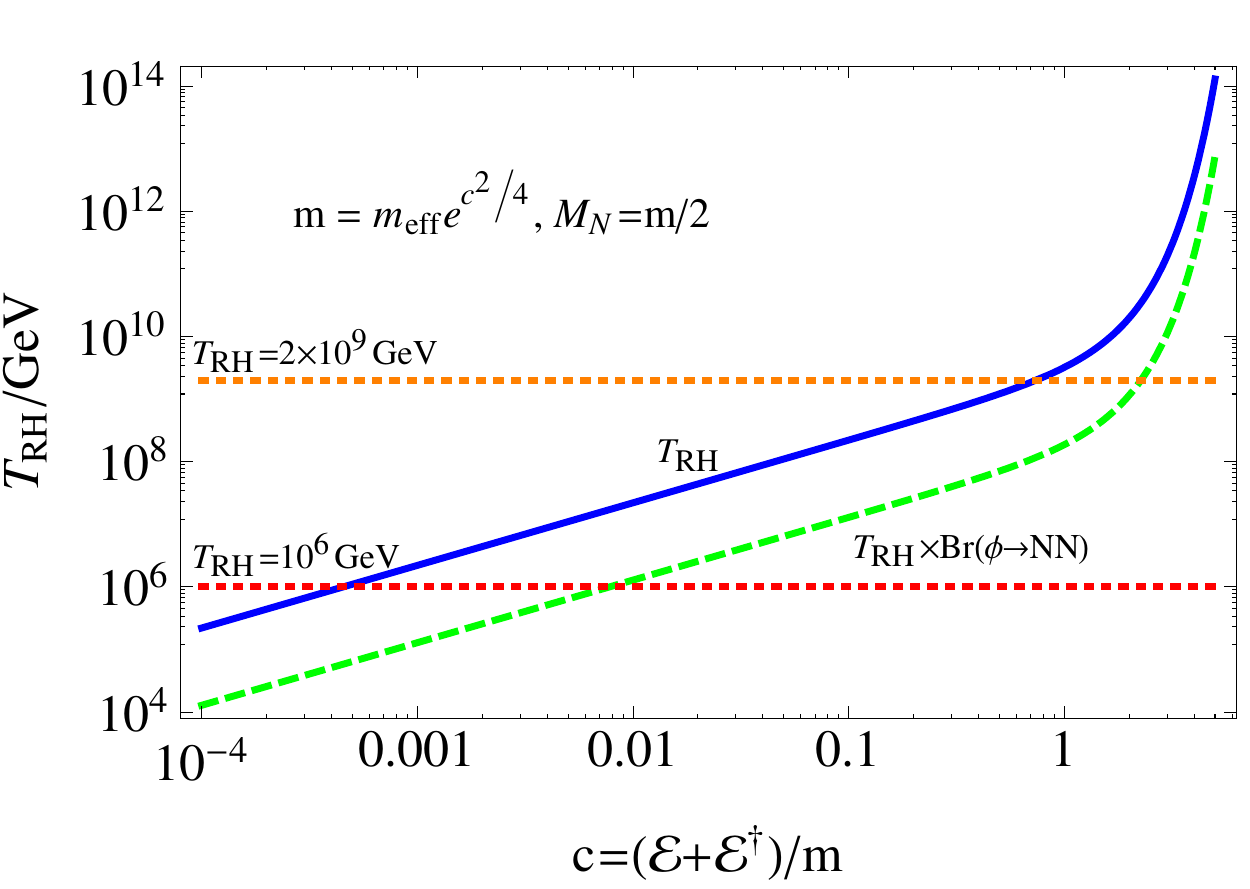}
 \end{center}
\caption{\sl \small
The real line shows the reheating temperature for a given parameter $c$.
The dashed line shows the reheating temperature times the branching ratio of the inflaton into the right-handed neutrino.
}
\label{fig:TRH}
\end{figure}

In Fig.~\ref{fig:TRH}, we show the relation between the reheating temperature $T_{\rm RH}\equiv 0.2 \sqrt{ \Gamma_{\rm tot}}$ and the parameter $c$, where $\Gamma_{\rm tot}$ is the total decay width of the inflaton.
Here, we assume $a_1=1$ and ${\cal E}$ is real.

Let us put a restriction on the reheating temperature, which is crucial for the discussion on the gravitino problem in the next section.
Throughout this paper, we assume that leptogenesis~\cite{Fukugita:1986hr} is responsible for the origin of the baryon asymmetry of the universe.
The thermal leptogenesis requires $T_{\rm RH}\gsim 2\times 10^9$ GeV~\cite{Giudice:2003jh,Buchmuller:2004nz}, and hence $c\gsim 0.7$.
Since the inflaton decays into the right-handed neutrino, non thermal leptogenesis~\cite{Kumekawa:1994gx,Lazarides:1999dm,Asaka:1999yd} is also possible.%
\footnote{Leptogenesis from inflaton decay is also discussed in Ref.~\cite{Lazarides:1991wu}, where the mechanism of generating the lepton asymmetry depends on the grand unification scale spectrum, however.}
In Fig.~\ref{fig:TRH}, we also show $T_{\rm RH} \times{\rm Br}(\phi\rightarrow NN)$ by a dashed line.
Here, it is assumed that $M_N = m/2$, so that the decay width of the inflaton into the right-handed neutrino is maximum.
Non thermal leptogenesis requires $T_{\rm RH} \times{\rm Br}(\phi\rightarrow NN)\times (2M_N / m) \gsim 10^6$ GeV~\cite{Lazarides:1999dm,Hamaguchi:2001gw}, and hence $c \gsim 0.008$.

In the following, we at least require $T_{\rm RH} \times{\rm Br}(\phi\rightarrow NN) \times (2M_N / m) > 10^{6}$ GeV, that is, $c>0.008$, so that non thermal leptogenesis is possible.
We also consider the more severe constraint from the successful thermal leptogenesis, $T_{\rm RH}>2\times 10^9$ GeV, that is, $c>0.7$.
This constraint should be satisfied when $M_N\ll m$ and hence ${\rm Br}(\phi\rightarrow NN) \times (2M_N / m)$ is suppressed.

\section{Gravitino problem and the gravitino mass}
\label{sec:gravitino}
It is known that gravitinos are in general produced through the decay of the inflaton, which results in the overproduction of the LSP~\cite{Kawasaki:2006gs,Asaka:2006bv,Dine:2006ii,Endo:2006tf,Endo:2006qk,Endo:2007ih}.
In this section, we first discuss how gravitinos are produced from the decay of the inflaton.
Then we discuss how large gravitino mass is required to avoid the LSP overproduction.

\subsection{Review on the decay of the inflaton into gravitinos}

Let us consider the simplest SUSY breaking model with the following (effective) superpotential,
\begin{eqnarray}
W = \mu^2 Z,
\end{eqnarray}
where $\mu^2 = \sqrt{3} m_{3/2}$ is the SUSY breaking scale and $Z$ is the SUSY breaking field.
Since the SUSY breaking field $Z$ does not obtain its mass from the superpotential, it should obtain its mass from the Kahler potential; otherwise, the SUSY breaking field obtains a large amplitude in the early universe, causing the cosmological Polonyi problem~\cite{Coughlan:1983ci,Ibe:2006am}.
The Kahler potential term which yields the mass term is
\begin{eqnarray}
\label{eq:Kahler mZ}
K = - \frac{1}{\Lambda^2} Z Z^\dag Z Z^\dag = -\frac{m_Z^2}{12 m_{3/2}^2} Z Z^\dag Z Z^\dag,
\end{eqnarray}
where $\Lambda \ll 1 $ is  some energy scale and $m_Z$ is the mass of the scalar component of $Z$.
This term is provided by interaction of the SUSY breaking field with other fields in the SUSY breaking sector.
The inflaton in general decays into those fields in the SUSY breaking sector, as is the case with MSSM fields.
Since the SUSY breaking sector fields couple to the SUSY breaking field $Z$, they eventually decay into the gravitino.
We examine this issue for concrete examples later.

The inflaton also decays into a pair of gravitinos through the mass mixing between the inflaton and the scalar component of the SUSY breaking field $Z$~\cite{Kawasaki:2006gs}.
For the Kahler and super potential in Eq.~(\ref{eq:Kahlersuper}), the mass mixing is given by
\begin{eqnarray}
V_{\rm mix} = \sqrt{3} c m_{3/2} m Z X^\dag + {\rm h.c.},
\end{eqnarray}
at around $Z = \Phi =X =0$.
The mixing angle between the inflaton and the SUSY breaking field is given by
\begin{eqnarray}
\label{eq:mixing angle}
\theta = \sqrt{\frac{3}{2}}c\frac{m_{3/2} m }{m_Z^2- m^2}.
\end{eqnarray}
The coupling between the scalar component of $Z$ and the gravitino, that is, the goldstino $\psi$ is given by the Kahler potential in Eq.~(\ref{eq:Kahler mZ}) as
\begin{eqnarray}
\label{eq:Zgoldstino}
{\cal L} = -\frac{\sqrt{3}}{6} \frac{m_Z^2}{m_{3/2}}Z^\dag \psi \psi + {\rm h.c.}.
\end{eqnarray}
From Eqs.~(\ref{eq:mixing angle}) and (\ref{eq:Zgoldstino}), we obtain the decay width of the inflaton into a pair of gravitinos,
\begin{eqnarray}
\label{eq:decay gravitino}
\Gamma(\phi \rightarrow 2 \psi_{3/2}) &=& \frac{c^2}{64\pi} m^3 \left(\frac{m_Z^2}{m_Z^2 - m^2}\right)^2 \nonumber\\
&\simeq&
\left\{
\begin{array}{ll}
\frac{c^2}{64\pi} m^3 & (m_Z \gg m)\\
\frac{c^2}{64\pi} \frac{m_Z^4}{m} & (m_Z \ll m).
\end{array}
\right.
\end{eqnarray}
The decay width is of the same order as that into MSSM fields if $m_Z \gg m$.

Now it is clear that the inflaton in general decays into gravitinos.
The gravitino eventually decays into the LSP.
The density parameter of the LSP is given by
\begin{eqnarray}
\label{eq:LSP abundance}
\Omega_{\rm LSP}h^2 \simeq \sum_f n_f {\rm Br}(\phi \rightarrow f ) \frac{3 T_{\rm RH}}{4 m} \frac{m_{\rm LSP}}{3.6\times 10^{-9}~{\rm GeV}}.
\end{eqnarray}
Here, $f$ denotes decay modes and $n_f$ is a number of gravitinos produced per the decay mode.
For example, $n_f=2$ for $f = 2 \psi_{3/2}$.

\subsection{Gravitino problem in strongly coupled SUSY breaking model}

We first discuss a strongly coupled SUSY breaking model.
To be concrete, let us consider the $SU(5)$ SUSY breaking model~\cite{Affleck:1983vc,Meurice:1984ai}.
The model is composed of an $SU(5)$ gauge theory with ${\bf 10}$ and ${\bf \bar{5}}$ representations.
Since there is no parameter expect for the gauge coupling,
$\Lambda$ and $\mu$ are as large as the dynamical scale of the $SU(5)$ gauge theory, $\Lambda_5$.
Assuming the naive dimensional analysis~\cite{Manohar:1983md,Luty:1997fk}, the Kahler and the super potentials are evaluated as
\begin{eqnarray}
W &=& c_1\frac{\Lambda^2_5}{4\pi}{\cal Z}, \nonumber\\
K &=& {\cal Z} {\cal Z}^\dag -  c_2\frac{16\pi^2}{\Lambda_5^2}{\cal Z} {\cal Z}^\dag {\cal Z} {\cal Z}^\dag.
\end{eqnarray}
where $c_1$ and $c_2$ are order one coefficients,
and ${\cal Z}$ is a composite field responsible for the SUSY breaking.
Here, we have assumed that only one composite field has a non-vanishing SUSY breaking $F$ term, for simplicity.

As shown in Eq.~(\ref{eq:decay anomaly}),
the inflaton decays into the $SU(5)$ gauge multiplet $V_5$ through the kinetic function if the dynamical scale is small enough, $m\gsim 2 \Lambda_5$.
Here, we assume that masses of hadrons of the $SU(5)$ gauge theory is as large as $\Lambda_5$.
The decay rate is given by
\begin{eqnarray}
\Gamma(\phi \rightarrow V_5V_5) = \frac{27 \alpha_5^2}{32\pi^2} c^2 m^3,
\end{eqnarray}
where $\alpha_5$ is the fine structure constant of the $SU(5)$ gauge theory.
Note that the decay rate is of the same order as the decay rate into MSSM particles (see Eqs.~(\ref{eq:decay treeMSSM}) and (\ref{eq:decay superMSSM})), and hidden hadrons eventually decay into gravitinos.

Even if the decay mode is kinematically closed, $m\lsim 2 \Lambda_5$, then the direct decay into gravitinos is unsuppressed since $m_{\cal Z} \sim \Lambda_5 \gsim m$ (see Eq.~(\ref{eq:decay gravitino})).
Thus, the decay of the inflaton inevitably produces gravitinos and the resultant density parameter of the LSP is
\begin{eqnarray}
\label{eq:SU5Omega}
\Omega_{\rm LSP}h^2 \simeq \frac{T_{\rm RH}}{m} \frac{m_{\rm LSP}}{3.6\times 10^{-9}~{\rm GeV}}.
\end{eqnarray}
The universe is over closed by the LSP unless
\begin{eqnarray}
m_{\rm LSP}\lsim 10~{\rm MeV} \frac{m}{1.5\times 10^{13}~{\rm GeV}}\frac{10^6~{\rm GeV}}{T_{\rm RH}}.
\end{eqnarray}
When $m_{\rm LSP}$ is such small, however, thermally produced LSPs over close the universe (recall the Lee-Weinberg bound~\cite{Lee:1977ua}) unless the LSP is degenerated with a charged SUSY particle.
Such a light charged SUSY particle is already excluded by various experiments.%
\footnote{
In the PGM, the photino LSP of a mass of $O(10)$ MeV is naturally obtained if $m_{3/2} = O(1) $ GeV. In this case, however, the electroweak symmetry breaking scale is also $O(1)$ GeV.
}
We will not consider the strongly coupled SUSY breaking model below.

\subsection{Gravitino problem in a SUSY breaking model with weak coupling}

The origin of the failure in the strongly coupled SUSY breaking model is that
either the decay of the inflaton into gravitinos or that into SUSY breaking sector fields is unsuppressed.
Note that simultaneous suppression of these two decay modes is achieved by realizing the following hierarchy,
\begin{eqnarray}
\label{eq:hierarchy}
 m_Z \ll m \ll m_{\rm SUSY-breaking},
\end{eqnarray}
where $m_{\rm SUSY-breaking}$ is the mass scale of SUSY breaking sector fields.
We show in this subsection that this hierarchy is easily achieved if the SUSY breaking sector involves weak couplings~\cite{Nakayama:2012hy}.

To be concrete, let us consider the IYIT SUSY breaking model~\cite{Izawa:1996pk,Intriligator:1996pu} with the $SU(2)$ gauge theory.
We introduce four fundamental representation of the $SU(2)$, $Q_i~(i = 1\mathchar`-4)$. Below the dynamical scale of the $SU(2)$, $\Lambda_2$,
the theory is described by meson fields with the deformed moduli constraint~\cite{Seiberg:1994bz},
\begin{eqnarray}
\label{eq:dynamical}
W_{\rm dyn} = 4\pi \Xi ({\rm Pf}M_{ij}- \frac{\Lambda_2^2}{16\pi^2}),
\end{eqnarray}
where $M_{ij} = - M_{ji}\sim Q_i Q_j/\Lambda_2$ are meson fields and $\Xi$ is a Lagrange multiplier field.
${\rm Pf}$ denotes the Pfaffian over indices $i,j$.
Here, we again assume the naive dimensional analysis and put order one coefficients to unity.
It can be seen that there are flat directions, in which ${\rm Pf} M_{ij} = \Lambda_2^2/ 16\pi^2$.

To fix the flat directions, let us introduce five singlet chiral multiplets, $Z_a (a = 1\mathchar`-5)$ and assume the following superpotential,
\begin{eqnarray}
\label{eq:tree}
W_{\rm tree} = \lambda c_{a,ij} Z_a Q_i Q_j, 
\end{eqnarray}
where $\lambda$ and $c_{a,ij}$ are constants.
To simplify our discussion,
we assume a global $SO(5)$ symmetry under which
$Z_a$ and $Q_i$ are the vector and the spinor representation of the $SO(5)$ symmetry. $c_{a,ij}$ should be appropriate Clebsh-Gordan coefficients.
Adding Eqs.~(\ref{eq:dynamical}) and (\ref{eq:tree}), we obtain the effective superpotential,
\begin{eqnarray}
\label{eq:eff}
W = \frac{\lambda}{4\pi} \Lambda_2 Z_aM_a + 4\pi \Xi ( M_a M_a + M^2- \frac{\Lambda_2^2}{16\pi^2}),
\end{eqnarray}
where we take linear combinations of meson fields and form a vector representation of the $SO(5)$, $M_a~(a = 1\mathchar`-5)$.
$M$ is the remaining independent linear combination.
Now, flat directions are fixed and the vacuum is given by $Z_a = M_a =0$, $M = \Lambda_2/4\pi$.

To break the SUSY, we add an additional singlet chiral multiplet $Z$ and add the superpotential,
\begin{eqnarray}
\label{eq:tree2}
\Delta W = y Z c_{ij}Q_i Q_j,
\end{eqnarray}
where $y$ is a constant and $c_{ij}$ is an appropriate Clebsh-Gordan coefficient to form a singlet of the $SO(5)$.
Adding Eqs.~(\ref{eq:eff}) and (\ref{eq:tree2}), we obtain the superpotential,
\begin{eqnarray}
\label{eq:eff super}
W =\frac{y}{4\pi} \Lambda_2 ZM + 
\frac{\lambda}{4\pi} \Lambda_2 Z_aM_a + 4\pi \Xi ( M_a M_a + M^2- \frac{\Lambda_2^2}{16\pi^2}).
\end{eqnarray}
Assuming $y\ll \lambda$, the vacuum is given by $Z_a \simeq M_a \simeq 0$, $M \simeq \Lambda_2/4\pi$.
The $F$ term of $Z$ is non-zero and hence the SUSY is spontaneously broken.

Let us discuss the decay of the inflaton into the SUSY breaking sector.
If the dynamical scale is small enough, $m\gsim 2 \Lambda_2$, the inflaton decays into gauge multiplets of the $SU(2)$,
as shown in Eq.~(\ref{eq:decay anomaly}).
The decay rate and $n_f$ are
\begin{eqnarray}
\label{eq:decay gauge}
\Gamma(\phi\rightarrow V_2 V_2) = \frac{3\alpha_2^2}{16\pi^3} c^2 m^3~~({\rm for}~m>2 \Lambda_2),~~n_{V_2V_2}\geq4,
\end{eqnarray}
where $\alpha_2$ is the fine structure constant of the $SU(2)$ gauge theory.
As in the case of the $SU(5)$ model, for $m > 2 \Lambda_2$, this decay mode is as dominant as decay modes into MSSM fields and hence the gravitino is overproduced.

If $m<2\Lambda_2$, on the other hand,
the mass of the inflaton is not far above the dynamical scale, and hence we can treat the decay of the inflaton into SUSY breaking sector fields by composite picture.
The decay rate through mass terms in Eq.~(\ref{eq:eff super}) and $n_f$ are
\begin{eqnarray}
\label{eq:decay mesons}
\Gamma(\phi \rightarrow Z_a M_a)&=&
\frac{5}{16\pi} c^2m \left(\frac{\lambda}{4\pi}\Lambda_2\right)^2
~~({\rm for}~m>2 \frac{\lambda}{4\pi}  \Lambda_2),~~
n_{Z_a M_a} =  4,\nonumber \\
\Gamma(\phi \rightarrow Z M) &=&
\frac{1}{16\pi} c^2m \left(\frac{y}{4\pi}\Lambda_2\right)^2
~~({\rm for}~m> \Lambda_2),~~
n_{Z M} =  4.
\end{eqnarray}
To be conservative, we assume that $\lambda \simeq 4\pi$. In this case, the decay into $Z_a M_a$ is kinematically forbidden. 

Now, we are at the point to show that the desired hierarchy in Eq.~(\ref{eq:hierarchy}) can be realized.
After integrating out $Z_a$, $M_a$ and $M$, we are left with the effective superpotential,
\begin{eqnarray}
\label{eq:super Z}
W_{\rm eff} = \frac{y}{16\pi^2} c_3 \Lambda_2^2 Z,
\end{eqnarray}
where $c_3=1$ is a constant, which we leave as a free parameter for later convenience.
The dynamical scale $\Lambda_2$ is related with the gravitino mass as
\begin{eqnarray}
\Lambda_2 = 3^{1/4}m_{3/2}^{1/2} 4\pi y^{-1/2} c_3^{-1/2} = 2.6\times 10^{12}~{\rm GeV}~y^{-1/2} \left(\frac{m_{3/2}}{10~{\rm TeV}}\right)^{1/2}c_3^{-1/2}.
\end{eqnarray}
The mass of the scalar component of $Z$ is given by the Kaher potential,
\begin{eqnarray}
K = - \frac{y^4}{16\pi^2\Lambda_2^2} Z Z^\dag Z Z^\dag,
\end{eqnarray}
and is as large as
\begin{eqnarray}
m_Z = \frac{2y^3}{(4\pi)^3} \Lambda_2 c_3 = 2.6\times 10^{9}~{\rm GeV} y^{5/2} \left(\frac{m_{3/2}}{10~{\rm TeV}}\right)^{1/2}c_3^{1/2}.
\end{eqnarray}
It can be seen that the hierarchy in Eq.~(\ref{eq:hierarchy}) is achieved for small $y$, and hence the overproduction of the LSP is avoided.

For a small $y$, however, the scalar component of $Z$ is light and the oscillation of the scalar $Z$ is induced in the early universe~\cite{Nakayama:2012hy}.
The oscillation eventually decays into gravitinos, which may lead to the overproduction of the LSP.
Let us estimate the abundance of the LSP from this contribution.
The potential of the scalar component of $Z$ during inflation is given by
\begin{eqnarray}
V(Z) = a_2 H_{\rm inf}^2 |Z|^2 + m_Z^2 |Z|^2 - (2 \sqrt{3}m_{3/2}^2 Z + {\rm h.c.}),
\end{eqnarray}
where $H_{\rm inf}$ is the Hubble scale during inflation and $a_2$ is an order one constant, which we assume to be positive.
Since $H_{\rm inf}\simeq 10^{14}$ GeV is larger than $m_Z$ for the parameter of interest, the Hubble induced mass term traps $Z$ to its origin during inflation. After inflation, as the Hubble scale drops below $m_Z$, $Z$ begins its oscillation around the origin,
\begin{eqnarray}
Z_0 = 2\sqrt{3}m_{3/2}^2/m_Z^2 = 1.2\times 10^8~{\rm GeV}~y^{-5}\frac{m_{3/2}}{10~{\rm TeV}}c_3^{-1},
\end{eqnarray}
with an initial amplitude $Z_i = Z_0$.
As anticipated, the amplitude is larger for smaller $y$.
The LSP abundance originated from the oscillation of $Z$ is given by
\begin{eqnarray}
\Omega_{\rm osc}h^2 = \frac{T_{\rm RH}}{4 m_Z} \frac{Z_i^2}{M_{\rm pl}^2}  \frac{m_{\rm LSP}}{3.6\times 10^{-9}~{\rm GeV}}.
\end{eqnarray}

\begin{figure}[tb]
\begin{center}
  \includegraphics[width=.5\linewidth]{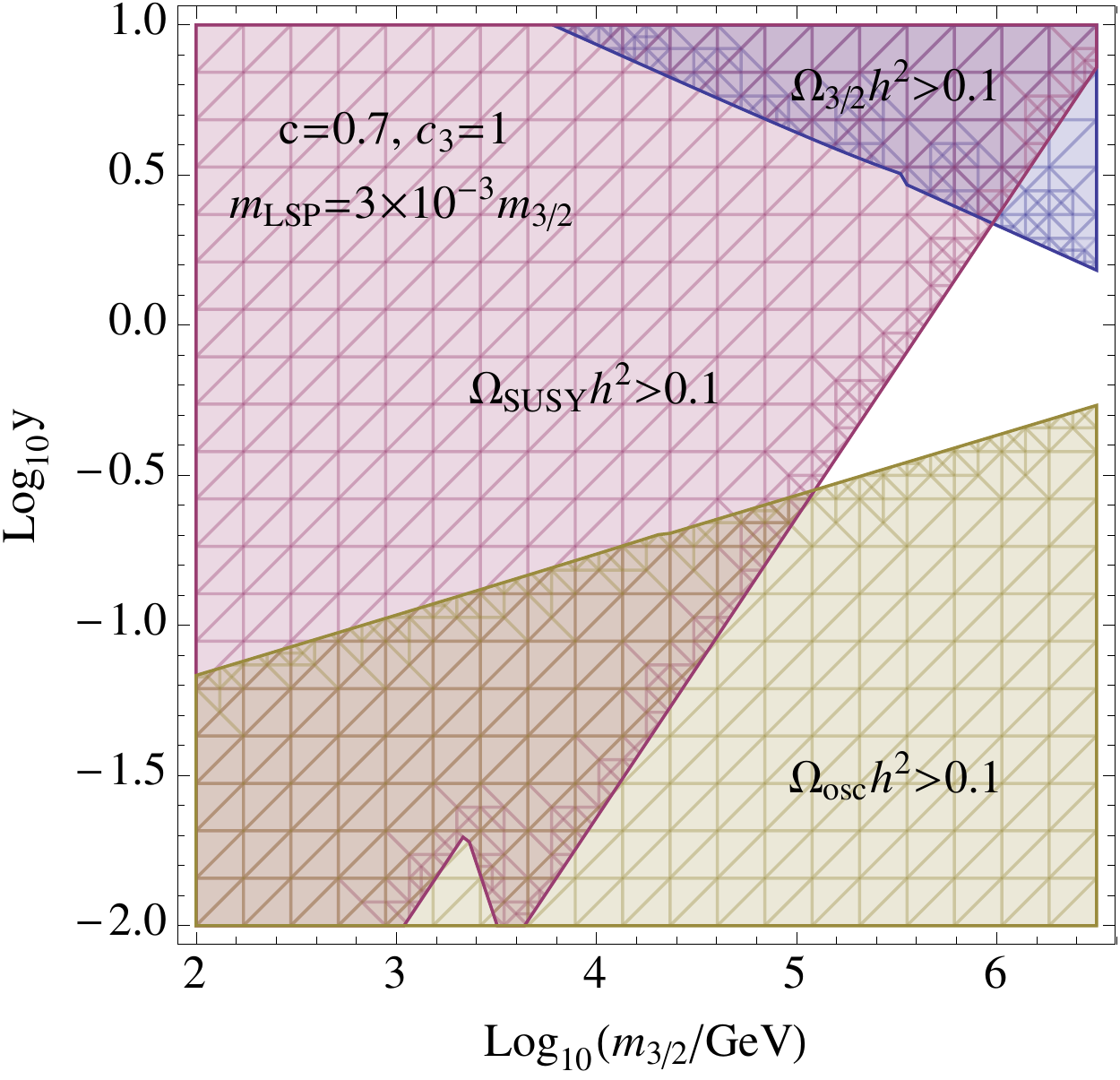}
 \end{center}
\caption{\sl \small
Constraint on the gravitino mass $m_{3/2}$ and the coupling of the SUSY breaking field $y$.
In the red-shaded region ($\Omega_{\rm SUSY}h^2 >0.1$),
the blue shaded region ($\Omega_{3/2}h^2 > 0.1$)
and the yellow shaded region ($\Omega_{\rm osc}h^2 >0.1$),
the universe is over closed by the LSP due to
the decay of the inflaton into SUSY breaking sector fields,
that into gravitino pairs,
and that of the SUSY breaking field into gravitinos, respectively.
We assume $c=0.7$, $c_3=1$ and $m_{\rm LSP}=3\times 10^{-3} m_{3/2}$.
}
\label{fig:m32th}
\begin{center}
  \includegraphics[width=.5\linewidth]{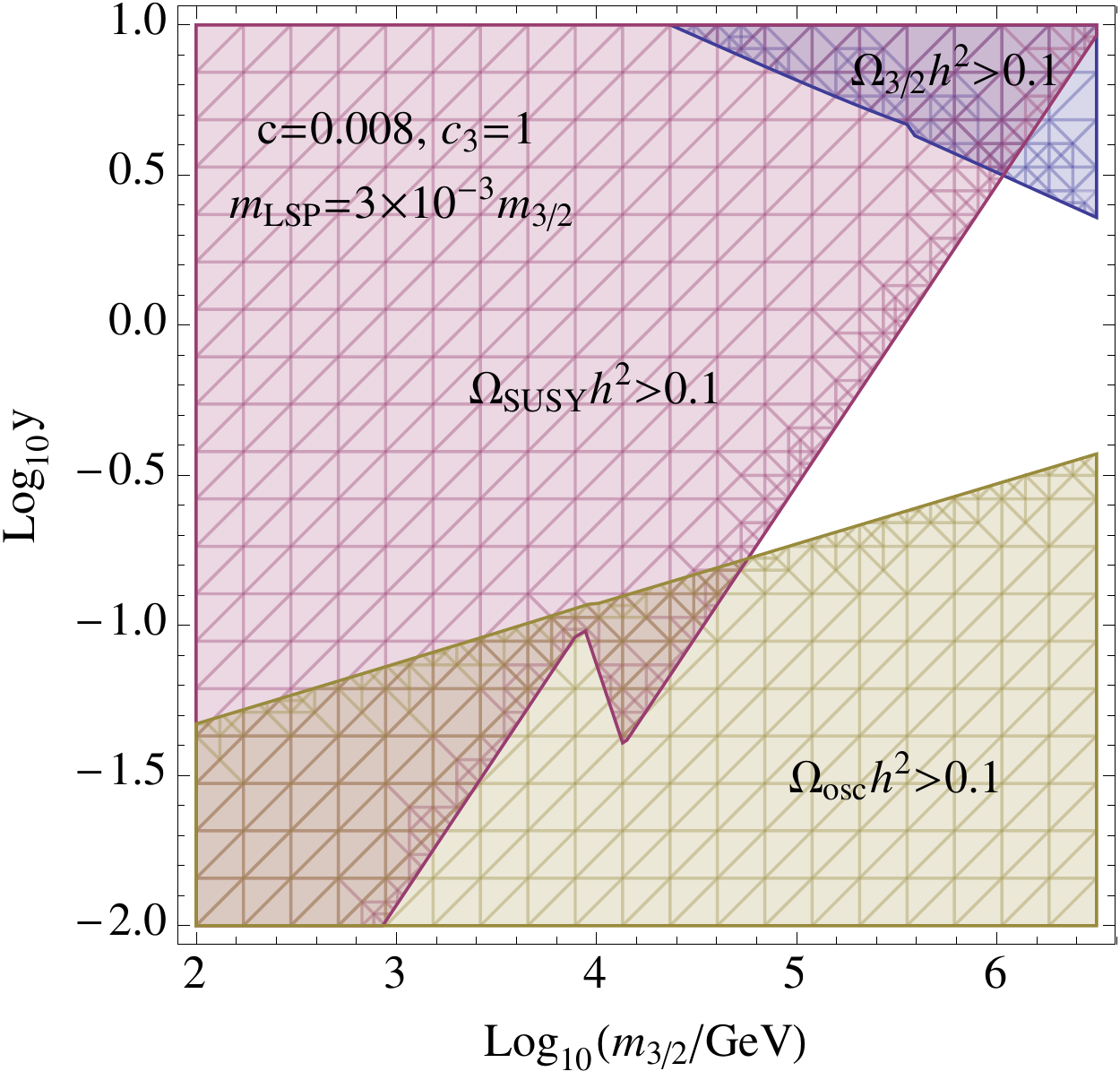}
 \end{center}
\caption{\sl \small
Same as Fig.~\ref{fig:m32th} but with $c=0.008$, $c_3=1$ and $m_{\rm LSP}=3\times 10^{-3} m_{3/2}$.
}
\label{fig:m32nt}
\end{figure}

Let us show how large gravitino mass is required.
In Fig.~\ref{fig:m32th},
we show the constraint on $m_{3/2}$ and $y$.
Here, we assume that $c=0.7$ (i.e. thermal leptogenesis is possible) and $m_{\rm LSP} = 3\times 10^{-3}m_{3/2}$.
In the red-shaded region ($\Omega_{\rm SUSY}h^2 >0.1$), the universe is over closed by the LSP due to the decay of the inflaton into SUSY breaking sector fields (see Eqs.~(\ref{eq:decay gauge}) and (\ref{eq:decay mesons})).
The right edge of this region is determined by the kinematical threshold, $m = \Lambda_2$.
In the blue shaded region ($\Omega_{3/2}h^2 > 0.1$), the decay of the inflaton into a pair of gravitinos causes the over closure.
In the yellow shaded region ($\Omega_{\rm osc}h^2 >0.1$), the oscillation of the SUSY breaking field leads to the over closure.
In Fig.~\ref{fig:m32nt}, we show the same constraint for $c=0.008$ (i.e. non thermal leptogenesis is possible).
From both figures, we see the constraint on the gravitino mass,%
\footnote{A similar conclusion is derived in Ref.~\cite{Nakayama:2014xca} where the BBN constraints are used.
Notice that we have obtained Eq.~(\ref{eq:constraint}) solely from constraints on the LSP DM density.
}
\begin{eqnarray}
\label{eq:constraint}
m_{3/2} > O(100)~{\rm TeV}.
\end{eqnarray}
It is remarkable that the constraint in Eq.~(\ref{eq:constraint}) coincides with what is expected in the PGM~\cite{Ibe:2006de,Ibe:2011aa,Ibe:2012hu}.

\begin{figure}[tb]
\begin{center}
  \includegraphics[width=.5\linewidth]{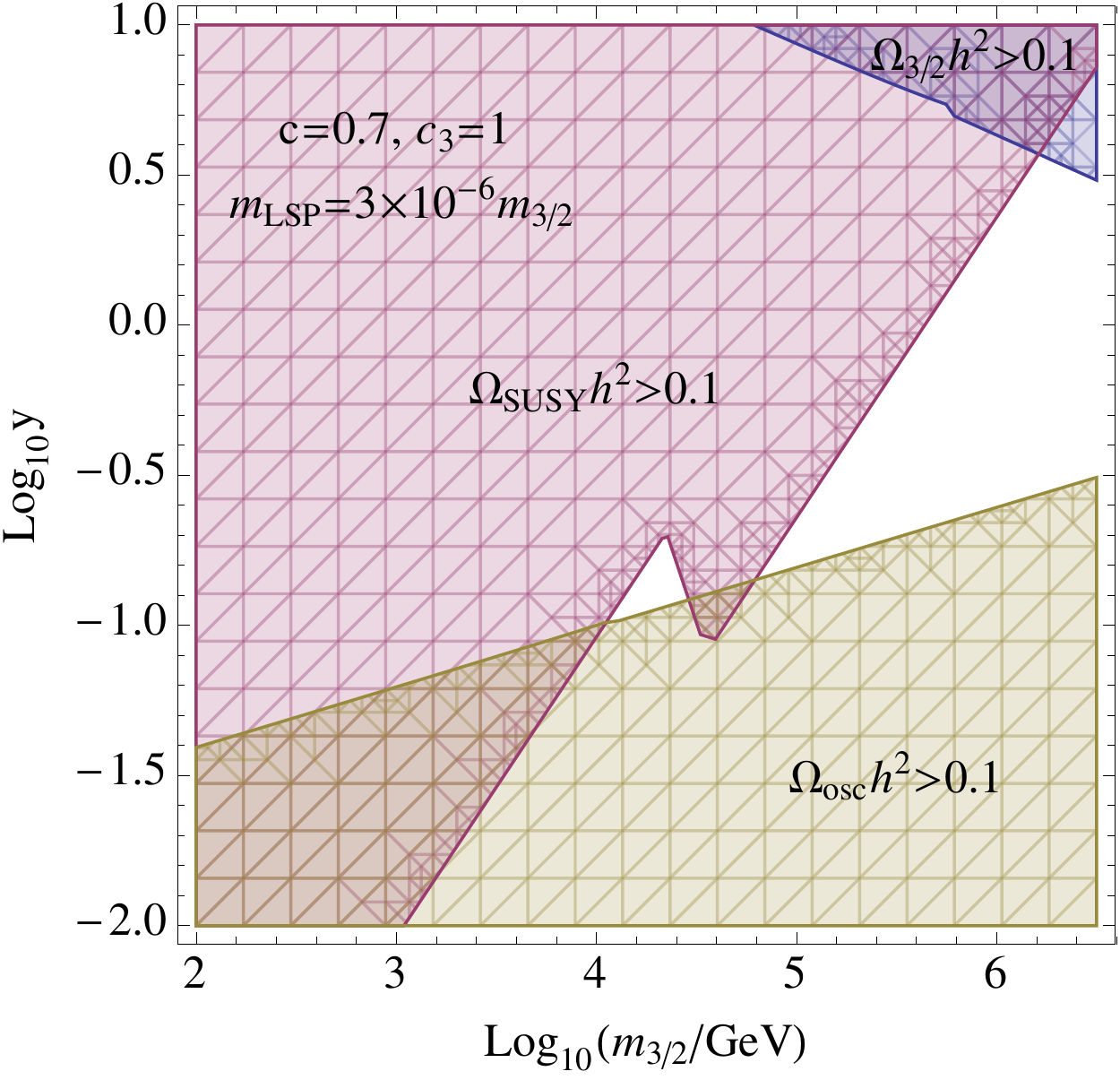}
 \end{center}
\caption{\sl \small
Same as Fig.~\ref{fig:m32th} but with $c=0.7$, $c_3=1$ and $m_{\rm LSP}=3\times10^{-6} m_{3/2}$.
}
\label{fig:m32thLSP}
\end{figure}

Let us discuss how we can avoid the constraint on the gravitino mass.
First, we have assumed that $m_{\rm LSP} = 3\times 10^{-3}m_{3/2}$ to obtain the constraint,
since it is determined by the anomaly mediation~\cite{Randall:1998uk,Giudice:1998xp}.
However, a lower mass for the LSP can be obtained by canceling the anomaly mediated contribution by the higgsino threshold correction~\cite{Giudice:1998xp}.
In Fig.~\ref{fig:m32thLSP}, we show the constraint on $m_{3/2}$ and $y$ for $(c,m_{\rm LSP}) = (0.7, 3\times 10^{-6}m_{3/2})$.
It can be seen that regions with $m_{3/2} = O(10)$~TeV is allowed.

Let us compare the plausibility of $m_{3/2} = O(10)$~TeV with that of $m_{3/2} = O(100)$~TeV in the landscape point of view.
Since we have no knowledge about distributions of parameters in the landscape, we discuss on our naive expectation in the following.
We note that different assumptions on the distribuions lead to different conclusions.

For the electroweak scale, $m_{3/2} = O(10)$~TeV would be more natural than $m_{3/2} = O(100)$~TeV 
by a factor of $(100~{\rm TeV})^2/ (10 {\rm TeV})^2 = 100$.
For the LSP mass, 
since the LSP mass is a complex parameter, $m_{\rm LSP}= 3\times 10^{-6} m_{3/2}$ would requires tuning of $(3\times 10^{-6}/3\times 10^{-3})^2\sim 10^{-6}$.
Thus, we naively expect that the region with $m_{3/2} = O(100)$~TeV may be more natural than the region with $m_{3/2} = O(10)$~TeV.

\begin{figure}[tb]
\begin{center}
  \includegraphics[width=.5\linewidth]{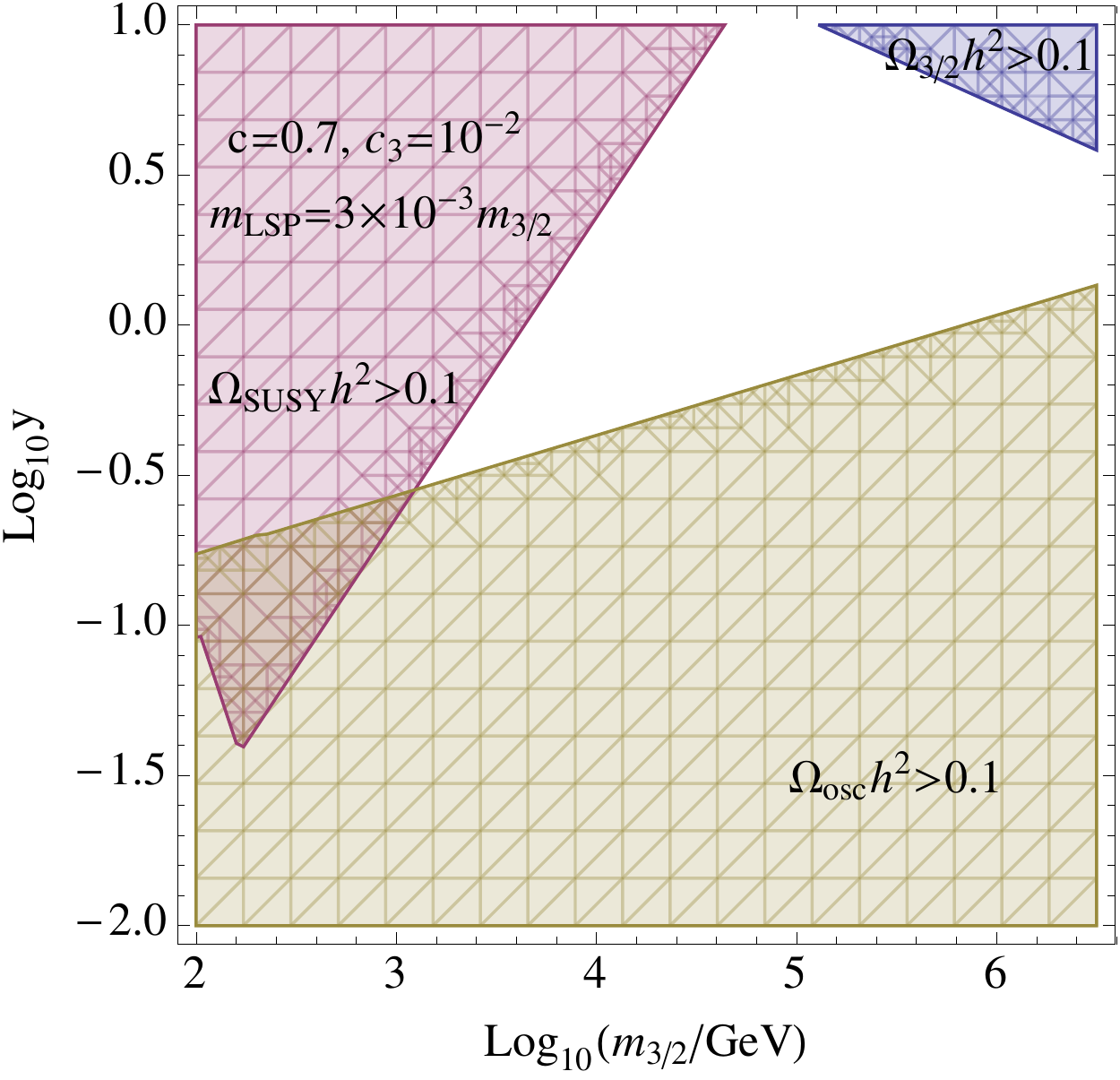}
 \end{center}
\caption{\sl \small
Same as Fig.~\ref{fig:m32th} but with $c=0.7$, $c_3=10^{-2}$ and $m_{\rm LSP}=3\times10^{-3} m_{3/2}$.
}
\label{fig:m32thSUSY}
\end{figure}

Second, we have assumed the $SO(5)$ symmetric IYIT model to simplify our discussion.
Without the $SO(5)$ symmetry, $c_3$ in Eq.~(\ref{eq:super Z}) is a constant which is determined by coupling constants in the SUSY breaking model.
If there is fine-tuned cancellation between condensation of hidden quarks which couple to the SUSY breaking field,
$c_3$ can be much smaller than $O(1)$. This cancellation further separates the SUSY breaking scale from the dynamical scale.
For given $m_{3/2}$ and $y$, the constraints shown in Figs.~\ref{fig:m32th} and \ref{fig:m32nt} are relaxed.
In Fig.~\ref{fig:m32thSUSY}, we show the constraint for $(c,c_3)=(1.4,10^{-2})$.
It can be seen that the region with $m_{3/2} = O(1)$~TeV survives.

Let us again naively compare the plausibility of $m_{3/2} = O(1)$~TeV with that of $m_{3/2} = O(100)$~TeV.
For the electroweak scale, $m_{3/2} = O(1)$~TeV would be more natural than $m_{3/2} = O(100)$~TeV 
by a factor of $(100~{\rm TeV})^2/ (1 {\rm TeV})^2 = 10^4$.
On the other hand, since $c_3$ is a complex parameter, $c_3=10^{-2}$ would require fine tuning of $10^{-4}$.
These two regions, $m_{3/2}= O(1)$ TeV and $O(100)$ TeV, may be equally plausible.


Third, we have assumed the minimal form of the Kahler potential.
By considering higher dimensional terms in the Kahler potential and tuning their coefficients, the decay of the inflaton into the SUSY breaking sector can be suppressed.
In principle, the gravitino mass of $O(100)$ GeV survives by the tuning.
However, to suppress all the decay modes, all the coefficients of the higher dimensional terms must be carefully chosen, which may require more fine tuning.


We should stress, finally, that all of the above arguments are merely a sketch on what kinds of fine tuning is needed to have the gravitino mass below $O(100)$ TeV.
Since we do not know distributions of relevant parameters in landscape, it is impossible for us to draw any definite conclusion on the most plausible gravitino mass.
However, the present analysis shows that it is not necessarily surprising that the nature has really chosen the gravitino mass of $O(100)$ TeV, even if the SUSY breaking scale is low energy biased in order to obtain the electroweak scale naturally.

\section{Discussion and conclusion}
In this paper, we have investigated compatibility of the PGM with chaotic inflation in supergravity.
We have shown that the inflaton should have a $Z_2$ odd parity to suppress the reheating temperature, avoiding the gravitino overproduction from thermal bath in the PGM.
We have also shown that the $Z_2$ symmetry is helpful for the inflaton to have consistent dynamics without tuning of parameters in the inflaton sector.

In order for the inflaton to decay, we assume that the $Z_2$ symmetry is broken by a small amount.
We have discussed the reheating process and the gravitino problem under the assumption of a small breaking of the $Z_2$ symmetry.
We have discussed how the gravitino overproduction by the decay of the inflaton can be avoided, and shown that the solution to the overproduction problem favors a gravitino mass far larger than the electroweak scale, $m_{3/2}\gsim O(100)$ TeV.

This consideration gives a new insight on the fine tuning problem in the high scale SUSY.
It is usually assumed that the gravitino mass of $O(100)$ GeV is natural,
for the electroweak scale is obtained without tuning of parameters in the MSSM.
It can be hardly understood why the nature chooses the gravitino mass of $O(100)$ TeV.
However, as we have shown in this paper, the gravitino mass of $O(100)$ GeV requires some amount of fine tuning to avoid the LSP overproduction.
Therefore, it may not be surprising even if the nature has chosen a high scale SUSY with the gravitino mass of $O(100)$ TeV.

In this paper, we have assumed the $Z_2$ symmetry to suppress the reheating temperature.
Another option is to assume the spacial separation of the inflaton sector and the MSSM sector in a higher dimensional theory.
Our discussion on the LSP overproduction is also applicable to this case.

We should note that we can replace the inflaton mass $m$ in Eq.~(\ref{eq:superKahler}) by a vacuum expectation value of some field.
Consider a $B-L$ gauge symmetry, for example, which is broken by a vacuum expectation value of a chiral multiplet $S$ with a $B-L$ charge of $+1$.
We assume that $X$ carries a $B-L$ charge of $-1$ so that the following superpotential is allowed~\cite{Buchmuller:2014rfa},
\begin{eqnarray}
W = k \Phi S X,~~k \vev{S} = m.
\end{eqnarray}
The Yukawa coupling $k$ represents a shift symmetry breaking.
We may take $k = O(0.1)$ and $\vev{S} = O(10^{-4})$ as an example.
The unwanted linear term $W = M^2 X$ is replaced by $W = {\cal M} \vev{S} X$, and the required condition $M^2 = {\cal M}\vev{S} \lsim m$ may be explained by ${\cal M} \lsim 0.1$ without the $Z_2$ symmetry.
 
In this paper, we have assumed that $m_{\rm LSP}=O(10^{-3}) m_{3/2}$, which is the case of the PGM.
In general gravity mediation models with a singlet SUSY breaking field (i.e. a Polonyi field), the LSP mass is expected to be of order the gravitino mass.
If it is this case, thermally produced LSPs will easily over close the universe unless the reheating temperature is far smaller than the LSP mass.
For $T_{\rm RH}\gsim 10^{6}$ GeV, the gravitino mass smaller than $10^{7}$ GeV is excluded.

Finally, let us comment on other inflation models.
The lower bound on the gravitino mass in Eq.~(\ref{eq:constraint}) is basically obtained from the condition that masses of SUSY breaking sector fields are larger than the inflaton mass.
Thus, for models with the inflation mass of $O(10^{13})$~GeV, a similar bound on the gravitino mass to Eq.~(\ref{eq:constraint}) will be obtained.
On the contrary, if models have the maximal reheating temperature,
the lower bound on the gravitino mass may be obtained~\cite{Harigaya:2012hn} so that enough LSPs are produced through the gravitino production in thermal bath to explain the DM density.

\section*{Acknowledgments}
We thank Brian Feldstein, Masahiro Ibe and Shigeki Matsumoto for fruitful discussion.
This work is supported by Grant-in-Aid for Scientific research from the
Ministry of Education, Science, Sports, and Culture (MEXT), Japan, No.~26287039 (T.T.Y.),
and also by World Premier International Research Center Initiative (WPI Initiative), MEXT, Japan (K.H. and T.T.Y.).
The work of K.H. is supported in part by a JSPS Research Fellowships for Young Scientists.

\end{document}